\documentclass[10pt, 5p]{elsarticle}

\usepackage[english]{babel}
\usepackage[utf8]{inputenc}
\usepackage{graphics}
\usepackage{graphicx}
\usepackage[x11names]{xcolor}
\usepackage{upgreek}
\usepackage{caption}
\usepackage{subcaption}
\usepackage{amsmath,amssymb,esint}
\usepackage{multirow}
\usepackage[colorlinks]{hyperref}

\biboptions{sort&compress,authoryear}

\journal{International Journal of Multiphase Flow}

\begin{document}

\begin{frontmatter}

\title{Water droplet dynamics and evaporation in airtanker firefighting}

\author{Fabian Denner\corref{cor1}}
\ead{fabian.denner@polymtl.ca}
\address{Department of Mechanical Engineering, Polytechnique Montréal, Montréal, H3T 1J4, Québec, Canada\vspace{-2.55em}}
\cortext[cor1]{Corresponding author: }

\begin{abstract}
This study presents the first systematic investigation of the dynamics of individual water droplets in the context of airtanker firefighting. While previous work has focused on ground-deposition patterns measured in standardized field tests, the droplet-scale mechanisms governing evaporation and transport have remained largely unexplored. A tailored model of the coupled momentum, heat, and mass transfer of an isolated water droplet in ambient air is proposed and applied to examine the evolution of droplets under a wide range of atmospheric conditions. The results demonstrate that droplet size governs the effectiveness of water delivery, the release height emerges as the dominant operational parameter, and relative humidity is the key atmospheric property. Increasing the release height lengthens the flight time and increases evaporative losses, while low relative humidity accelerates evaporation, particularly for droplets smaller than one millimeter. Only droplets within a narrow range of initial radii, $150\,\upmu\mathrm{m} \lesssim r_{\mathrm{d},0} \lesssim 3\,\mathrm{mm}$, are able to reach the ground following an airtanker release, with smaller droplets fully evaporating during their fall and larger droplets being subject to secondary atomization. Although airtanker releases involve very large liquid volumes and complex spray dynamics, the present analysis deliberately isolates droplet-scale behavior and does not resolve collective spray effects, wake interactions, or turbulence. The findings therefore serve as a physically consistent baseline for droplet evaporation and transport, forming a foundation for spray-resolved modeling efforts aimed at improving airtanker delivery strategies.
\end{abstract}

\begin{keyword}
Water droplets \sep Droplet dynamics \sep Evaporation \sep Humidity \sep Airtanker firefighting\\~\\[-0.5em]
\textcopyright~2026. This manuscript version is made available under the CC-BY 4.0 license. \href{http://creativecommons.org/licenses/by/4.0/}{http://creativecommons.org/licenses/by/4.0/}
\end{keyword}

\end{frontmatter}

\section{Introduction}

\begin{figure*}[t]
	\includegraphics[width=0.48\linewidth]{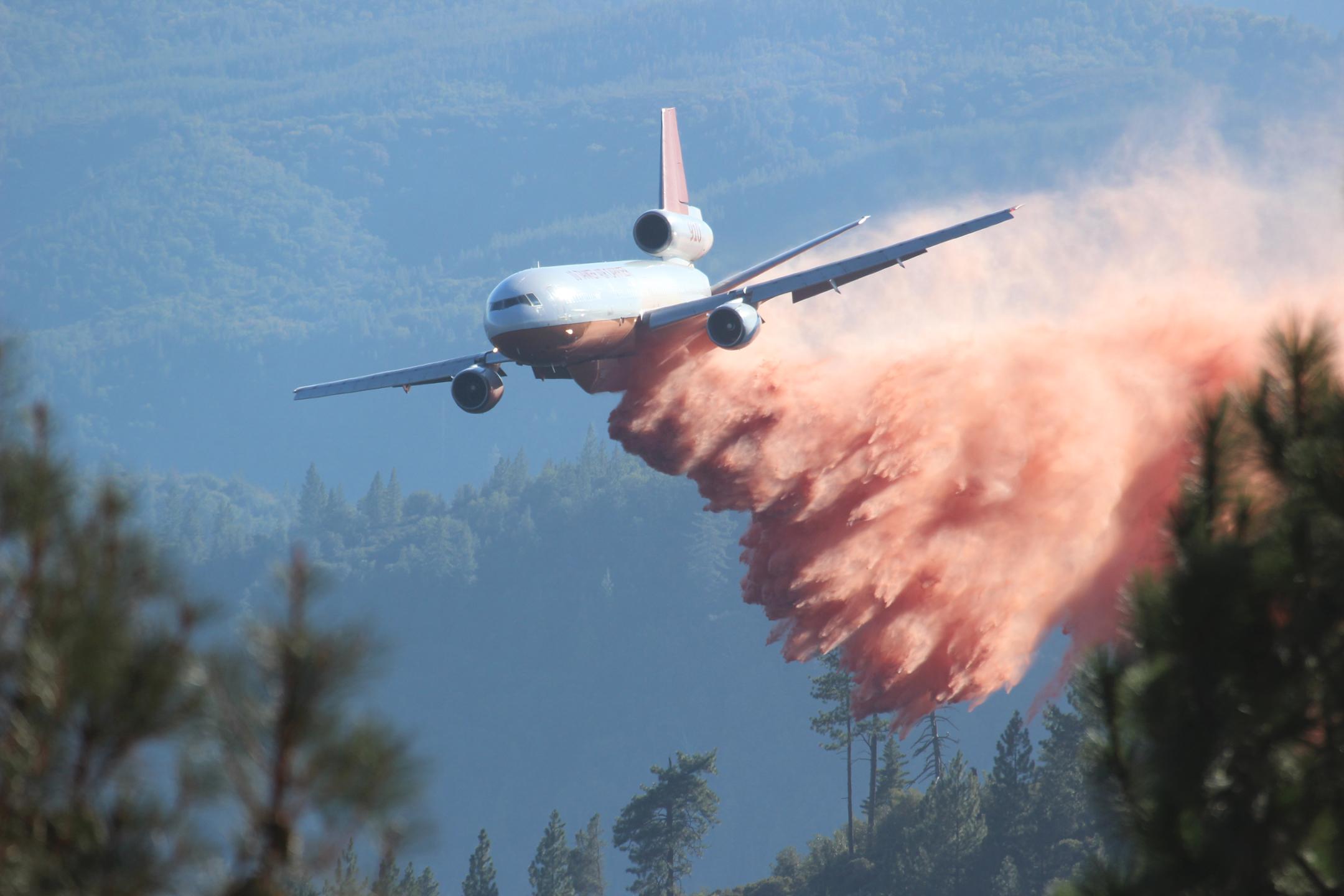}\hfill
	\includegraphics[width=0.48\linewidth]{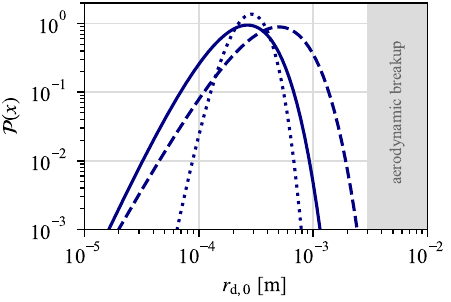}
	\caption{(a) A McDonnell Douglas DC-10-30 airtanker releases fire retardant. U.S. Forest Service photo by Mike McMillan (``20130817-FS-UNK-0024'' by U.S. Department of Agriculture, CC BY 2.0). (b) Droplet size distributions produced by the primary atomization process in airtanker firefighting defined by Eq.~\eqref{eq:gamma}, as proposed by \citet{Legendre2024}. The solid line shows the probability density related to ligament breakup, the dashed and dotted lines show the probability density related to bag breakup.}
	\label{fig:plane-DSD}
\end{figure*}

With extreme wildfires becoming more than twice as frequent and intense in the last two decades \citep{Cunningham2024}, the fluid dynamics governing aircraft-delivered water and fire retardants to suppress wildfires demand renewed scrutiny. While amphibious airtankers, such as the emblematic Canadair CL-415, can deliver up to $6000$ liters of water sourced directly from a local river, lake or sea, the largest airtanker currently in operation, the McDonnell Douglas DC-10-30 shown in Figure \ref{fig:plane-DSD}(a), can deliver over $35000$ liters of water or retardant in a single release. 
The release of such a large volume of liquid at velocities of up to $70 \, \text{m/s}$ during airtanker firefighting results in a complex fragmentation and distribution process that includes primary atomization driven by surface instabilities, droplet transport, coalescence and breakup, heat and mass transfer. Targeted atomization and deposition of the released liquid is critical for the success of wildfire suppression using airtankers, yet our understanding of these fluid dynamic processes at the scales relevant for airtanker firefighting is still in its infancy \citep{Legendre2024}.

In airtanker firefighting two different operational modes are common \citep{Plucinski2019, Struminska2024, Legendre2024}. A \textit{direct attack} involves releasing water directly onto the fire front in the initial stages of a wildfire, with the aim of cooling and containing the fire while it still burns with low intensity. 
In contrast, during an \textit{indirect attack}, fire retardant is released ahead of the fire front to establish fire lines that guide or quench the fire. 
The testing of aircraft-based water and retardant delivery systems typically takes place above an open field on which measurement cups are arranged on a grid, the so-called \textit{cup-and-grid method}, to measure the distribution of liquid delivered by an airtanker \citep{Gu2023, Suter2000, Legendre2014}. No fire or additional heat source are usually present in these tests. Although test data suggests that even during these tests up to 45\% of water evaporates \citep{Legendre2014}, key information influencing evaporation and, hence, the liquid volume arriving on the ground, such as the ambient temperature and relative humidity, are \textit{not} taken into account when analyzing the effectiveness of airtanker delivery systems \citep{Gu2023}.
Similarly, the coupled momentum, heat and mass transfer of droplets in airtanker firefighting  has not been studied systematically to date \citep{Gu2023} and  existing models of airtanker firefighting \citep{Amorim2011,Amorim2011a,Sun2024a} do not take heat and mass transfer into account.

After exiting an airtanker, the liquid undergoes primary atomization dominated by a Kelvin-Helmholtz instability, with subsequent bag and ligament breakup \citep{Legendre2024, Sun2024a, Christensen2025}. The droplet size distribution produced by this atomization process is not known precisely and recent research efforts have been focusing on addressing this question \citep{Rouaix2023, Calbrix2023, Christensen2025}.
Theoretical considerations by \citet{Legendre2024} suggest that droplets with a radius of $20 \, \upmu\text{m} \lesssim r_{\text{d},0} \lesssim 2 \, \text{mm}$ are produced, described by the gamma distributions associated with the different breakup mechanisms typical for liquid atomization processes under high shear \citep{Villermaux2007}, shown in Figure \ref{fig:plane-DSD}(b). As a result of the acting aerodynamic forces, the largest stable water droplets falling freely in air have a radius of $r_\text{d} \approx 2.5-3 \, \text{mm}$ \citep{Hobbs2004, Villermaux2009, Pruppacher2010, Loftus2021, Zhao2023c}.\footnote{Larger water droplets with a radius of up to $5 \, \text{mm}$ have been observed in clouds but are unstable in free fall \citep{Hobbs2004, Zhao2023c}.} 
Consequently, \citet{Legendre2024} concluded that the droplets formed by the primary atomization process are unlikely to undergo secondary atomization. After formation, the droplets are in free fall as part of a dense spray exhibiting considerable local variations in temperature and relative humidity that govern the heat and mass transfer between the droplets and their surroundings, determining ultimately how much liquid reaches the fire or the ground. 

The atmospheric conditions in which water or retardant is delivered has an important influence on the effectiveness of firefighting efforts, directly influencing the heat and mass transfer between the liquid and its surrounding, as well as the displacement of the liquid by wind drift. In particular for the dynamics of water droplets, the meteorological literature provides ample evidence of the importance of relative humidity and temperature in determining which droplets ``survive'' a fall and which droplets disappear as a result of evaporation even under normal atmospheric conditions \citep{Pruppacher2010,Loftus2021}. In the vicinity of a wildfire, however, the atmospheric conditions are more complex, with hot plumes leading to elevated air velocities above the fire that can grow several kilometers in height \citep{Lareau2017} and reach air temperatures above $100 \, ^\circ \text{C}$ at heights of tens of meters above the fire \citep{Clements2010}. The orientation and vertical air velocity of the plumes are, in turn, strongly influenced by the canopy configuration and prevailing wind speeds \citep{Cervantes2025}, further complicating the environment in which airtanker firefighting takes place. In addition, airtanker operations themselves can change the local atmospheric conditions. Field tests by \citet{Wheatley2023} have demonstrated that the large amount of water delivered by an airtanker may locally raise the relative humidity by approximately $15 \, \%$-points and decrease the temperature by an average $3 \, ^\circ\text{C}$ in forested areas.

Previous studies on aerial firefighting have  focused on large-scale delivery characteristics and have not explicitly considered the fate of individual droplets, which ultimately governs both the effectiveness of water delivery and the interpretation of standardized field tests used to assess airtanker systems. To address this gap, the present study develops a coupled momentum, heat, and mass transfer model for a single water droplet in air and applies it to quantify the influence of release height, ambient temperature, and relative humidity on droplet motion, evaporation, and the effective liquid volume reaching the ground. Although airtanker releases involve very large liquid volumes and complex spray dynamics, the fate of this liquid is determined at the droplet scale. Accordingly, the analysis deliberately isolates single-droplet behavior to establish a physically consistent baseline for evaporation and transport under conditions relevant to airtanker firefighting.

In the following, Section \ref{sec:method} proposes a model to describe the coupled momentum, heat, and mass transfer of a single water droplet. The droplet dynamics under different atmospheric conditions are studied in Section \ref{sec:results} and the evaporated liquid volume resulting from an airtanker water release is analyzed in Section \ref{sec:evpoarated-volume}. To put these findings into perspective, a representative case of a field test is considered in Section \ref{sec:cases}. The findings of this study are summarized and conclusions are drawn in Section \ref{sec:conclusions}.

\section{Droplet model}
\label{sec:method}

A single water droplet falling freely in a uniform air atmosphere at ambient pressure $p_\infty = 101325 \, \text{Pa}$ is considered in this study in order to isolate the fundamental transport mechanisms governing droplet evaporation and motion during its fall. Given a typical release height of $100 \, \text{m}$ or less, the assumption of a uniform atmosphere is justified \citep{Minzner1977}. Radiative heat transfer is neglected to clearly isolate the effects of relative humidity and ambient temperature, which are particularly important in tests used to evaluate the performance of airtanker delivery systems, in which no fire or other significant additional heat sources are present. Sections \ref{sec:droplet-motion} and \ref{sec:droplet-evaporation} propose a tailored model for the gravity-driven motion as well as heat and mass transfer of a single water droplet in an air atmosphere, Section \ref{sec:fluidproperties} describes the relevant fluid properties, and Section \ref{sec:validation} provides a validation of the model.

\subsection{Droplet motion}
\label{sec:droplet-motion}

The velocity $\mathbf{u}_\text{d}$ and position $\mathbf{x}_\text{d}$ of a droplet with radius $r_\text{d}$ readily follow from Newton's second law as
\begin{align}
	\frac{\text{d}\mathbf{u}_\text{d}}{\text{d}t} &= 
    \frac{\mathbf{F}_g + \mathbf{F}_{\text{D}}}{\rho_\text{w} V_\text{d}}\\
	\frac{\text{d}\mathbf{x}_\text{d}}{\text{d}t} &= \mathbf{u}_\text{d},
\end{align}
where $\rho_\text{w}$ is the water density, $V_\text{d} = 4\pi r_\text{d}^3/3$ is the droplet volume, $\textbf{F}_g$ is the force due to gravity, and $\textbf{F}_\text{D}$ is the drag force acting on the droplet.
Considering that the motion of a water droplet in air under typical atmospheric conditions is dominated by drag and gravity forces, unsteady forces are neglected.

The force acting on the droplet due to gravity is given as
\begin{equation}
    \mathbf{F}_{g} = \mathbf{g} \left(\rho_\text{w} - \rho_\text{a}\right) V_\text{d}
\end{equation}
where $\mathbf{g} = (0 , 0, -9.81)^\text{T} \, \text{m/s}^2$ is the gravitational acceleration.
The drag force acting on the droplet is defined as
\begin{equation}
     \mathbf{F}_{\text{D}} = \frac{\rho_\text{a}}{2} C_\text{D} \pi r_\text{d}^2 \, | \mathbf{u}_\text{a} - \mathbf{u}_\text{d} | \, (\mathbf{u}_\text{a} - \mathbf{u}_\text{d}),
\end{equation}
where $\rho_\text{a}$ is the air density and 
$\mathbf{u}_\text{a}$ is the 
air velocity. Because the droplet may deform during its fall, following the recent work of \citet{Zhao2025b}, the drag coefficient of the droplet is defined as \citep{Liu1993b}
\begin{equation}
	C_\text{D} = C_\text{D,0} \left[ 1 + 2.632 \left(1 - \frac{r_\text{d}^2}{r_\text{max}^2} \right) \right],
\end{equation}
where the drag coefficient of a spherical droplet with the same volume is \citep{Clift1971}
\begin{equation}
	C_\text{D,0} = \frac{24}{\text{Re}} \, \left(1 + 0.15 \, \text{Re}^{0.687} + \frac{0.0175 \, \text{Re}}{1 + 42500 \, \text{Re}^{-1.16}} \right)
\end{equation}
and the maximum radius of the deformed droplet is \citep{Hsiang1992}
\begin{equation}
	r_\text{max} = r_\text{d} \left(1 + 0.19 \, \sqrt{\text{We}}\right).
\end{equation}
The Reynolds and Weber numbers of the droplet are
\begin{align}
	\text{Re} &= \frac{2  r_\text{d} \rho_\text{a} |\mathbf{u}_\text{a}-\mathbf{u}_\text{d}|}{\mu_\text{a}} \label{eq:Re} \\
	\text{We} &= \frac{2 r_\text{d} \rho_\text{a} (\mathbf{u}_\text{a}-\mathbf{u}_\text{d})^2}{\sigma},
\end{align}
respectively,
where $\mu_\text{a}$ is the dynamic viscosity of air and $\sigma$ is the surface tension coefficient of the air-water interface of the droplet. 

The maximum possible size of a droplet falling freely is determined by its stability under the acting aerodynamic forces. Following the work of \citet{Zhao2023c}, who considered the gravity-enhanced Rayleigh-Taylor instability of a freely falling droplet, the critical Weber number for droplet breakup is taken to be
\begin{equation}
	\text{We}_\text{crit} = \frac{4}{3 C_\text{D}} \left[4 \pi^2 \left(\frac{r_\text{d}}{r_\text{max}}\right)^4 - \text{Bo} \left(\frac{r_\text{d}}{r_\text{max}}\right)^2\right],
\end{equation}
where the Bond number of the droplet is given as
\begin{equation}
	\text{Bo} = \frac{\rho_\text{w} g \left(2 r_\text{d}\right)^2 }{\sigma}.
\end{equation}
A droplet is assumed to break up if $\text{We} > \text{We}_\text{crit}$.

\subsection{Heat and mass transfer}
\label{sec:droplet-evaporation}

The change of mass $m_\text{d} = \rho_\text{w} V_\text{d}$ of a water droplet in an air atmosphere is described by \citep{Rogers1996}
\begin{equation}
	\frac{\text{d}m_\text{d}}{\text{d}t} = \rho_\text{w} 4 \pi r_\text{d}^2 \frac{\text{d}r_\text{d}}{\text{d}t}= 8 \pi r_\text{d}^2 h_m M_\text{w} (n_\infty - n_\text{s}),
\end{equation}
where 
\begin{equation}
	h_m = \frac{\text{Sh}\,D_\text{w,a}}{2 r_\text{d}}
\end{equation}
is the mass transfer coefficient, $\text{Sh}$ is the Sherwood number, $D_\text{w,a}$ is the diffusion coefficient of water vapor in air, $M_\text{w}$ is the molar mass of water, and $n$ is the molar concentration of water. Based on the ideal gas law, $n = p R /T$, where $R=8.3145 \, \text{J/(K mol)}$ is the universal gas constant and $T$ denotes the absolute temperature, the molar concentration in the far field is
\begin{equation}
	n_\infty = \phi_\infty \, \frac{p_\text{sat}(T_\infty)}{R \, T_\infty},
\end{equation}
where $\phi_\infty$ is the ambient relative humidity and $T_\infty$ is the ambient temperature of the air in the far field, and the molar concentration at the droplet surface is
\begin{equation}
	n_\text{s} = \frac{p_\text{sat}(T_\text{d})}{R \, T_\text{d}} ,
\end{equation}
where $p_\text{sat}$ is the saturation pressure of water and $T_\text{d}$ denotes the droplet temperature. Consequently, the vapor pressure in air is obtained from the saturation pressure at the ambient temperature $T_\infty$ and relative humidity $\phi_\infty$, and the air layer in contact with the droplet is assumed to be saturated, such that the partial pressure of the vapor at the droplet surface equals the saturation pressure at the droplet temperature. Including these definitions for the molar concentrations, the temporal change of the droplet radius is described by
\begin{align}
	\frac{\text{d}r_\text{d}}{\text{d}t} &= \text{Sh} \, \frac{D_\text{w,a} M_\text{w}}{2 r_\text{d} \, \rho_\text{w} \, R} \left(\phi_\infty \,  \frac{p_\text{sat}(T_\infty)}{T_\infty} - \frac{p_\text{sat}(T_\text{d})}{T_\text{d}} \right). \label{eq:drdt}
\end{align}

\begin{table*}
	\centering
	\caption{Coefficients for the simplified polynomial description, Eq.~\eqref{eq:air_poly}, of the properties of air based on the work of \citet{Tsilingiris2018}.}\label{table:gasproperties}
	\begin{tabular}{l|ccccc}
		 Property & $a_1$ & $b_1$ & $b_2$  & $c_1$ & $c_2$\\ \hline 
		$\rho_\text{a} \ \left[{\text{kg}}/{\text{m}^3}\right]$ & $1.2932$ & $-4.6688 \times 10^{-3}$ & $-8.6965 \times 10^{-4}$ & $1.6165 \times 10^{-5}$ & $2.2437 \times 10^{-5}$
		\\
		$\mu_\text{a} \ \left[\text{Pa s}\right]$ & $1.7192 \times 10^{-5}$ & $4.9467 \times 10^{-8}$ & $-2.2425 \times 10^{-9}$ & $-4.2412 \times 10^{-11}$ & $-3.2389 \times 10^{-10}$
		\\
		$k_\text{a} \ \left[{\text{W}}/{\text{K m}}\right]$ & $2.4031 \times 10^{-2}$ & $7.5904 \times 10^{-5}$ & $-3.1199 \times 10^{-6}$ & $-4.8653 \times 10^{-8}$ & $-1.3015 \times 10^{-7}$
		\\
		$c_{p,\text{a}} \ \left[{\text{J}}/{\text{kg K}}\right]$ & $1006.4$ & $3.9654 \times 10^{-2}$ & $2.0101$ & $4.15208 \times 10^{-4}$ & $-1.673 \times 10^{-1}$\\
		\hline
	\end{tabular}
\end{table*}

Assuming the temperature of the water droplet is spatially uniform and radiative heat transfer is negligible, the energy equation of the droplet is given as
\begin{align}
	c_{p,\text{w}} m_\text{d} \frac{\text{d}T_\text{d}}{\text{d}t} = - \frac{\text{d}Q}{\text{d}t} + L_\text{w} \frac{\text{d}m_\text{d}}{\text{d}t}, \label{eq:dedt}
\end{align}
where $c_{p,\text{w}}$ and ${L}_\text{w}$ are the specific isobaric heat capacity and the latent heat of vaporization of water, respectively. The convective heat flux between the droplet and its surrounding follows from Newton's law of cooling as
\begin{equation}
	\frac{\text{d}Q}{\text{d}t} = h_T S_\text{d} \left(T_\text{d}-T_\infty\right), \label{eq:dQdt}
\end{equation}
where $S_\text{d}=4\pi r_\text{d}^2$ is the surface area of the droplet. The heat transfer coefficient $h_\text{T}$ is expressed as
\begin{equation}
	h_T = \frac{\text{Nu}\,k_\text{a}}{2 r_\text{d}},
\end{equation}
where $k_\text{a}$ is the thermal conductivity of air and $\text{Nu}$ is the Nusselt number. Inserting Eq.~\eqref{eq:dQdt} into Eq.~\eqref{eq:dedt} and rearranging for the temperature derivative yields 
\begin{align}
	\frac{\text{d}T_\text{d}}{\text{d}t} &= \frac{3}{r_\text{d} \, c_{p,\text{w}}} \left( \text{Nu} \, k_\text{a} \, \frac{T_\infty-T_\text{d}}{2 r_\text{d} \, \rho_\text{w}} + L_\text{w} \, \frac{\text{d}r_\text{d}}{\text{d}t} \right). \label{eq:dTdt}
\end{align}
An equivalently formulated heat and mass transfer model (Eqs.~\eqref{eq:drdt} and \eqref{eq:dTdt}) has previously been applied, for instance, to falling rain droplets \citep{Loftus2021}. The assumption of a spatially uniform droplet temperature is justified by the small Biot number $\text{Bi} = h_T r_\text{d} / k_\text{w} \ll 1$, where $k_\text{w}$ is the thermal conductivity of water, for the considered water droplets with a radius of $r_\text{d} \lesssim 3$ mm in air, indicating negligible internal temperature gradients.

To account for the enhanced mass transfer of a moving droplet, the Sherwood number of water droplets moving in an air atmosphere is defined as \citep{Beard1971, Pruppacher1979}
\begin{align}
	\text{Sh} = \begin{cases}
		2 + 0.316 \left({\text{Re}} \, \text{Sc}^{2/3}\right), & \text{Re}^{1/2} \, \text{Sc}^{1/3} < 1.4 \\
		1.56 + 0.616 \left(\text{Re}^{1/2} \, \text{Sc}^{1/3}\right), & \text{Re}^{1/2} \, \text{Sc}^{1/3} \geq 1.4 
	\end{cases} \label{eq:Sh}
\end{align}
where $\text{Sc} = \mu_\text{a}/(\rho_\text{a} D_\text{w,a})$ is the Schmidt number. Assuming an equivalence between heat and mass transfer, the Nusselt number $\text{Nu}$ is typically defined in the same manner as the Sherwood number, simply replacing the Schmidt number by the Prandtl number. However, with this approach the heat and mass transfer model presented in Eqs.~\eqref{eq:drdt} and \eqref{eq:dTdt} yields terminal droplet temperatures below the wet-bulb temperature $T_\text{wb}$, leading to an underpredicted evaporation rate. Previous studies have frequently made use of specific definitions of the mixture properties, the Sherwood or the Nusselt numbers to correct the heat and mass transfer of evaporating droplets in various environments \citep{Ashgriz2011,Smolik2001,Sazhin2006,Mahmud2016}. 

In contrast, the clear focus of the current study on water droplets in ambient air allows to leverage high-accuracy correlations of the wet-bulb temperature in the studied conditions. To this end, the Nusselt number is defined as 
\begin{equation}
	\text{Nu} = \text{Nu}_0 + \max\{0,\Delta \text{Nu}_\text{wb}\},
\end{equation}
where $\Delta \text{Nu}_\text{wb}$ is a correction based on the wet-bulb temperature $T_\text{wb}$.
The base value of the Nusselt number is obtained using the same correlation as for the Sherwood number,
\begin{align}
	\text{Nu}_0 = \begin{cases}
		2 + 0.316 \left({\text{Re}} \, \text{Pr}^{2/3}\right), & \text{Re}^{1/2} \, \text{Pr}^{1/3} < 1.4 \\
		1.56 + 0.616 \left(\text{Re}^{1/2} \, \text{Pr}^{1/3}\right), & \text{Re}^{1/2} \, \text{Pr}^{1/3} \geq 1.4 
	\end{cases} \label{eq:Nu}
\end{align}
with the Prandlt number defined as
\begin{align}
	\text{Pr} &= \frac{\mu_\text{a} c_{p,\text{a}}}{k_\text{a}},
\end{align}
where $c_{p,\text{a}}$ is the specific isobaric heat capacity of air.
The correction to the Nusselt number follows from Eq.~\eqref{eq:dTdt} under the assumptions $T_\text{d} = T_\text{wb}$ and $\text{d}T_\text{d}/\text{d}t = 0$ as
\begin{equation}
	\Delta \text{Nu}_\text{wb} =  \frac{2 r_\text{d} \rho_\text{w} L_\text{w}}{k_\text{a} \left(T_\text{wb}-T_\infty\right)}\, \frac{\text{d}r_\text{d}}{\text{d}t} - \text{Nu}_0, \label{eq:dNu}
\end{equation}
ensuring the correct asymptotic balance of the sensible and latent heat fluxes, such that the droplet can assume the wet-bulb temperature associated with the ambient conditions.
The wet-bulb temperature is calculated by the empirical expression of \citet{Stull2011},
\begin{align}
	T_\text{wb} &= 273.15 \, \text{K} + \Theta_\infty \arctan \!\left(0.151977 \sqrt{\phi_\infty^\star+8.313659} \right) \nonumber \\ &+ \arctan \left(\Theta_\infty + \phi_\infty^\star \right)
	- \arctan \left(\phi_\infty^\star - 1.676331 \right) \nonumber \\ &+ 0.00391838 \left(\phi_\infty^\star\right)^{1.5} \arctan\left( 0.023101 \phi_\infty^\star\right)-4.686035, \label{eq:Twb}
\end{align}
where $\phi_\infty^\star = \phi_\infty \cdot 100 \, \%$ and $\Theta_\infty$ is the ambient temperature in units $^\circ\text{C}$. 
Eq.~\eqref{eq:dNu} does not imply that the droplet temperature assumes the wet-bulb temperature instantaneously. Instead, it enforces thermodynamic consistency of sensible and latent heat fluxes by ensuring that the wet-bulb temperature is recovered as the correct asymptotic equilibrium state. The droplet temperature therefore evolves dynamically toward this equilibrium, as described by Eq.~\eqref{eq:dTdt}.

This formulation of the coupled heat and mass transfer for a single droplet is intended to capture droplet-scale behavior and does not attempt to represent collective effects such as vapor accumulation or droplet-droplet interactions within a dense spray.

\subsection{Fluid properties}
\label{sec:fluidproperties}

The density $\rho_\text{a}$, the dynamic viscosity $\mu_\text{a}$, the thermal conductivity $k_\text{a}$, and the specific isobaric heat capacity $c_{p,\text{a}}$ of air are defined, following the work of \citet{Tsilingiris2018}, based on the ambient temperature $\Theta_\infty$ (in units $^\circ\text{C}$) and relative humidity $\phi_\infty$ using simplified polynomial correlations of the form 
\begin{equation}
	\psi(\Theta_\infty,\phi_\infty) = a_1 + \Theta_\infty (b_1 + b_2 \, \phi_\infty) + \Theta_\infty^2 (c_1 + c_2 \, \phi_\infty), \label{eq:air_poly}
\end{equation}
with the coefficients used for $\psi \in \{\rho_\text{a}, \mu_\text{a}, k_\text{a}, c_{p,\text{a}}\}$ given in Table \ref{table:gasproperties}.

\begin{figure*}[t]
	\centering
	\includegraphics[width=0.95\linewidth]{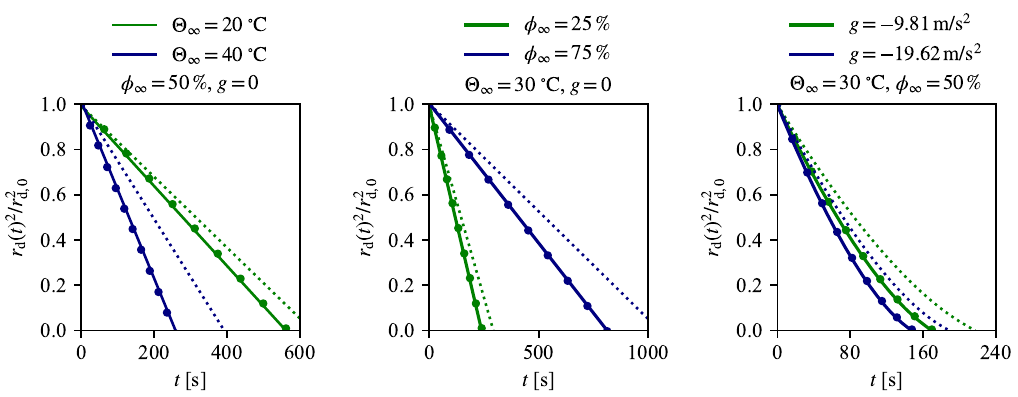}
	\caption{Evolution of the droplet surface area of static ($g=0$) or freely falling ($g<0$) water droplets with an initial radius of $r_\text{d,0} = 300 \, \upmu\text{m}$ in air under different ambient conditions. The solutions provided by the heat and mass transfer model presented in Section \ref{sec:droplet-evaporation} with the Nusselt number correction proposed in Eq.~\eqref{eq:dNu} are shown by the solid lines, the solutions obtained without the Nusselt number correction are shown by the dotted lines. The circular markers represent the reference solution obtained by solving Eq.~\eqref{eq:r2}.}
	\label{fig:validation_r2}
\end{figure*}

\begin{figure*}[t]
	\centering
	\includegraphics[width=0.95\linewidth]{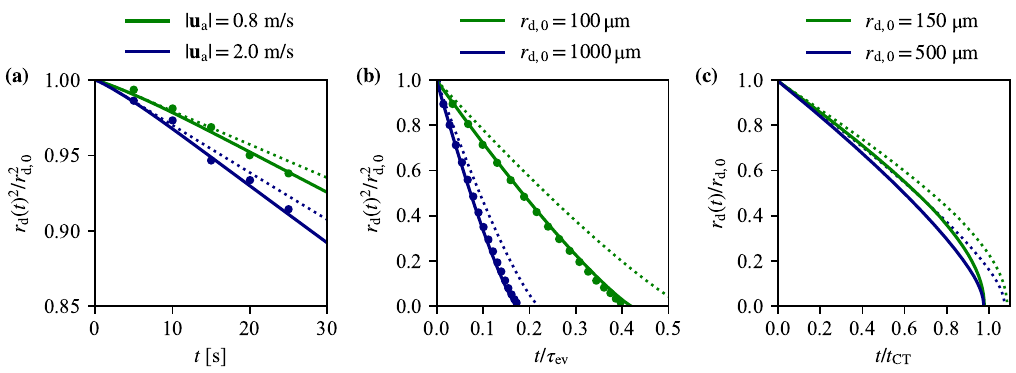}
\caption{Evolution of the relative droplet surface area and relative droplet radius. The solutions provided by the heat and mass transfer model presented in Section \ref{sec:droplet-evaporation} with the Nusselt number correction proposed in Eq.~\eqref{eq:dNu} are shown by the solid lines, the solutions obtained without the Nusselt number correction are shown by the dotted lines. (a) A water droplet with initial radius $r_\text{d,0} = 600 \, \upmu\text{m}$ and initial temperature $\Theta_\text{d,0} = 15 \, ^\circ\text{C}$ in an air flow with velocity $|\mathbf{u}_\text{a}|\in \{0.8,2\} \, \text{m/s}$, temperature $\Theta_\infty = 30 \, ^\circ\text{C}$ and relative humidity $\phi_\infty = 0.3$. The circular markers show the experimental measurements of \citet{Fujita2010}. (b) A water droplet with initial radii $r_\text{d,0} \in \{100,1000\} \, \upmu\text{m}$ and initial temperature $\Theta_\text{d,0} = 24 \, ^\circ\text{C}$ in an air flow with velocity $|\mathbf{u}_\text{a}|= 5 \, \text{m/s}$, temperature $\Theta_\infty=30\, ^\circ\text{C}$ and relative humidity $\phi_\infty = 0.5$. The circular markers show the reference results of \citet{Woo2011} and the time is scaled with the characteristic evaporation time $\tau_\text{ev}$ given in Eq.~\eqref{eq:tauev}. (c) A water droplet with initial radii $r_\text{d,0} \in \{150,500\} \, \upmu \text{m}$ and initial temperature $\Theta_\text{d,0} = 34 \, ^\circ\text{C}$ falling freely in air with temperature $\Theta_\infty = 20 \, ^\circ\text{C}$ and relative humidity $\phi_\infty = 0.5$. The time is scaled with the corresponding evaporation time $t_\text{CT}$ reported by \citet{Cavazzuti2023}.}
	\label{fig:validation_Wh}
\end{figure*}

Water has a molecular weight of $M_\text{w} = 18.0153 \, \text{g/mol}$, a surface tension coefficient defined as \citep{Vargaftik1983}
\begin{equation}
	\sigma(T_\text{d}) = 0.2358 \, \frac{\text{N}}{\text{m}} \, \hat{T}_\text{c,w}^{1.256} \, (1-0.625 \, \hat{T}_\text{c,w}),
\end{equation}
and its latent heat of vaporization is given by the correlation \citep{Xiang1997}
\begin{equation}
	L_\text{w}(T_\text{d}) = \left(6.8477 \, \dfrac{\hat{T}_\text{d}^{0.325} + 1.7383 \, \hat{T}_\text{c,w}^{0.835}}{1 + 1.3913 \, \hat{T}_\text{c,w}} \right)  \frac{R\,  T_\text{c,w}}{M_\text{w}},
\end{equation}
where the reduced temperatures are defined as
\begin{align*}
	\hat{T}_\text{d} &= \dfrac{T_\text{c,w}-T_\text{d}}{T_\text{d}}\\
	\hat{T}_\text{c,w} &= \dfrac{T_\text{c,w}-T_\text{d}}{T_\text{c,w}},
\end{align*}
and $T_\text{c,w}=647.096 \, \text{K}$ is the critical temperature of water. The density and specific isobaric heat capacity of water exhibit only small variations over the considered temperature range ($\approx 1 \, \%$ and $\approx 0.1 \, \%$, respectively) and are, thus, assumed to be constant with $\rho_\text{w}= 997 \, \text{kg/m}^3$ and $c_{p,\text{w}}= 4181 \, \text{J/(kg K)}$ \citep{Wagner2002}.

The saturation vapor pressure of water is calculated using the correlation \citep{Huang2018}
\begin{equation}
	p_\text{sat}(T) = \exp \left(34.494 - \frac{4924.99}{T-36.05}\right) \left(T -168.15 \right)^{-1.57},
\end{equation}
where $p_\text{sat}$ has the unit Pa,
and the diffusion coefficient of water vapor in air at ambient pressure $p_\infty = 101325 \, \text{Pa}$ is given by \citep{Massman1998}
\begin{equation}
	D_\text{w,a}(T_\infty) = 2.178 \, \times 10^{-5} \, \frac{\text{m}^2}{\text{s}} \left(\frac{T_\infty}{273.15 \, \text{K}}\right)^{1.81}. 
\end{equation}

\subsection{Validation}
\label{sec:validation}

Since dedicated experimental measurements of individual droplet behavior under conditions relevant for airtanker water delivery are not available, the droplet model is validated using theoretical considerations as well as suitable single-droplet test cases from the literature.

\citet{Langmuir1918} first reported a linear reduction of the droplet surface area during quasi-steady, diffusion-limited evaporation from a spherical droplet to the surrounding gas, now commonly referred to as the $d^{2}$-law. Because it yields a closed-form solution with a well-defined slope, the $d^{2}$-law provides a clear benchmark against which numerical models of the interfacial mass flux can be validated. Considering the temporal change of the droplet radius defined by Eq.~\eqref{eq:drdt} with the wet-bulb temperature as the terminal droplet temperature \citep{Cavazzuti2023,Woo2011}, the $d^2$-law representing the change of the relative droplet surface area is approximated as
\begin{equation}
	\frac{r_\text{d}^2(t)}{r_\text{d,0}^2} \approx 1 - t \, \frac{\text{Sh}\, D_\text{w,a} M_\text{w}}{r_\text{d,0}^2 \, \rho_\text{w} \, R} \left(\frac{p_\text{sat}(T_\text{wb})}{T_\text{wb}} - \phi_\infty \,  \frac{p_\text{sat}(T_\infty)}{T_\infty}  \right). \label{eq:r2}
\end{equation}
The corresponding characteristic evaporation time for a stationary droplet follows with $\text{Sh} = 2$ as
\begin{equation}
	\tau_\text{ev} = \frac{r_\text{d,0}^2 \, \rho_\text{w} \, R}{2\, D_\text{w,a} M_\text{w}} \left(\frac{p_\text{sat}(T_\text{wb})}{T_\text{wb}} - \phi_\infty \,  \frac{p_\text{sat}(T_\infty)}{T_\infty}  \right)^{-1}, \label{eq:tauev}
\end{equation}
which can be computed \textit{a priori}.
Figure \ref{fig:validation_r2} shows the results of the evolution of the relative  surface area of a droplet with $r_\text{d,0} = 300 \, \upmu\text{m}$ for different representative cases. The heat and mass transfer model presented in Section \ref{sec:droplet-evaporation} including the correction of the Nusselt number proposed in Eq.~\eqref{eq:dNu}, shown by the solid lines in Figure \ref{fig:validation_r2}, produces results that are in excellent agreement with the reference solution provided by Eq.~\eqref{eq:r2}, shown as circular markers in Figure \ref{fig:validation_r2}. In contrast, without the proposed correction of the Nusselt number ($\Delta \text{Nu}_\text{wb} = 0$), see the dotted lines in Figure \ref{fig:validation_r2}, the evaporation rate is systematically underpredicted and the droplet lifetime is, hence, overpredicted.

A similar picture emerges when comparing the proposed droplet model to results published previously in the literature, for droplets with initial radii in the range $100 \, \upmu\text{m} \leq r_\text{d,0} \leq 1000 \, \upmu\text{m}$ falling freely in air or in an air flow for representative ambient conditions, shown in Figure \ref{fig:validation_Wh}. The presented droplet model, including the correction of the Nusselt number on the basis of the wet-bulb temperature, provides a consistent and accurate prediction of the droplet evaporation, whereas the evaporation rate is underpredicted without the Nusselt number correction. The comparisons shown in Figure \ref{fig:validation_Wh} further demonstrate that the coupled model for momentum, heat and mass transfer presented in Sections \ref{sec:droplet-motion} and \ref{sec:droplet-evaporation}, with the fluid properties given in Section \ref{sec:fluidproperties}, is capable of accurately predicting the fate of isolated droplets in the size range and in conditions relevant for airtanker firefighting.

\section{Droplet dynamics}
\label{sec:results}

The dynamic behavior of water droplets in airtanker firefighting is subject to a complex environment that governs the motion and evaporation of each individual droplet. 
Large airtankers used in direct attacks release their liquid cargo from a height of $30-100 \, \text{m}$ at a velocity of $40 - 70 \, \text{m/s}$ \citep{Legendre2014, Legendre2024}\footnote{Small airtractors primarily built for agricultural crop dusting can deliver water and retardants from heights of only several meters at speeds below $20\, \text{m/s}$.}, after which the liquid undergoes a rapid atomization process. 
The Kelvin-Helmholtz instability dominates the atomization of the continuous liquid volume released by the airtanker, producing liquid ligaments and bags that subsequently break up into droplets \citep{Legendre2024}. 
The work of \citet{Legendre2024} suggests that the primary atomization process produces droplets with radii in the range $20 \, \upmu\text{m} \lesssim r_{\text{d},0} \lesssim 2 \, \text{mm}$, see Figure \ref{fig:plane-DSD}(b). 

\subsection{Droplet motion}

For the purpose of this study, the droplets are assumed to fall freely under the action of gravity and without additional driving forces, such that their terminal velocity is determined solely by the prevailing balance of drag and gravity forces. The initial droplet temperature is $\Theta_\text{d,0} = 25 \, ^\circ\text{C}$, unless mentioned otherwise.

\begin{figure}[t]
	\centering
	\includegraphics[width=\linewidth]{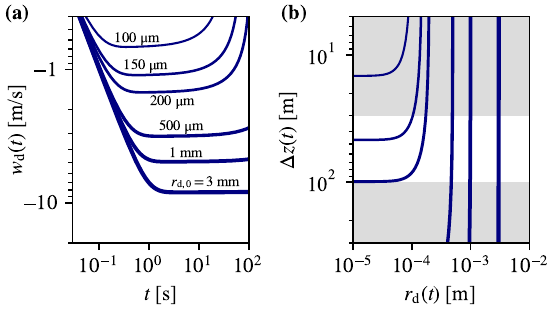}
	\caption{Evolution of (a) the vertical velocity $w_\text{d}$ as a function of time and (b) the falling distance $\Delta z$ as a function of the radius $r_\text{d}$,  of water droplets falling freely in a quiescent air atmosphere with temperature $\Theta_\infty = 25 \, ^\circ\text{C}$ and relative humidity $\phi_\infty = 0.5$. The droplets have an initial temperature of $\Theta_\text{d,0}=25 \, ^\circ\text{C}$ and initial radii of $r_\text{d,0} \in \{0.1, 0.15, 0.2, 0.5, 1, 3\} \, \text{mm}$, with thicker lines corresponding to larger radii. The typical release height of $30-100 \, \text{m}$ in airtanker firefighting is explicitly highlighted in (b).}
	\label{fig:wrz}
\end{figure}

Figure \ref{fig:wrz}(a) shows the vertical velocity of water droplets with an initial radius in the range $100 \, \upmu\text{m} \leq r_\text{d,0} \leq 3 \, \text{mm}$ falling freely in a quiescent air atmosphere with temperature $\Theta_\infty = 25 \, ^\circ\text{C}$ and relative humidity $\phi_\infty =0.5$. Under these ambient conditions, the maximum stable initial radius of the droplets is predicted to be $3.11 \, \text{mm}$, which is in good agreement with previous studies \citep{Zhao2023c,Villermaux2009}. While millimeter-sized droplets fall at a (nearly) constant terminal velocity, smaller droplets slow down considerably in the course of their fall as their size reduces due to evaporation. To better understand what this means for airtanker firefighting, Figure \ref{fig:wrz}(b) shows the radius of these droplets as a function of the distance they have fallen. For the typical release height in airtanker firefighting of $30 -100 \, \text{m}$ (assuming the liquid falls a negligible distance during primary atomization), only droplets with an initial radius of approximately $200 \, \upmu\text{m}$ and larger have a realistic chance of arriving on the ground. Large droplets with an initial radius of $r_\text{d,0} > 1 \, \text{mm}$ are not affected strongly by the ambient conditions. Considering that the majority of droplets is expected to have an initial radius of $\mathcal{O} (100 \, \upmu\text{m})$, see Figure \ref{fig:plane-DSD}(b), evaporation is likely an important factor in the effectiveness of airtanker firefighting.

\begin{figure}[t]
	\includegraphics[width=\linewidth]{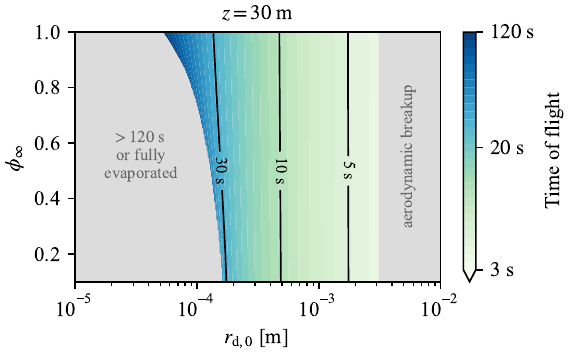}
	\\
	\includegraphics[width=\linewidth]{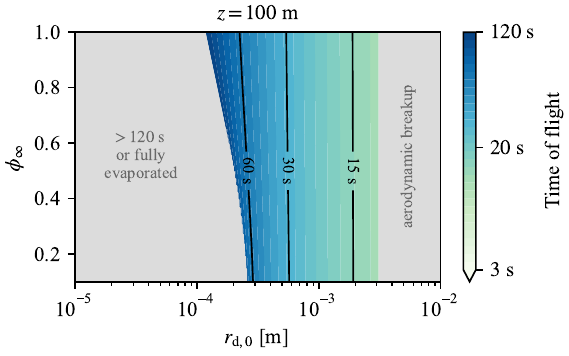}
	\caption{Phase maps of the time of flight of water droplets falling freely in an air atmosphere with a temperature of $\Theta_\infty = 25 \, ^\circ\text{C}$ for a fall  from $z \in \{30,100\} \, \text{m}$, as a function of the initial droplet radius $r_\text{d,0}$ and the relative humidity $\phi_\infty$. The initial droplet temperature is $\Theta_\text{d,0} = 25 \, ^\circ\text{C}$.}
	\label{fig:falltime}
\end{figure}

A further difficulty when releasing large amounts of water from an airplane are the prevailing wind conditions. Figure \ref{fig:falltime} shows phase maps of the time of flight of droplets falling freely from heights of $z \in \{30,100\} \, \text{m}$, in function of their initial radius $r_\text{d,0}$ and the relative humidity $\phi_\infty$. Millimeter-sized droplets reach the ground in a matter of seconds, whereas smaller droplets fall for many tens of seconds or evaporate fully before reaching the ground. As a result, small droplets ($r_\text{d,0} < 1 \, \text{mm}$) may be subject to considerable wind drift. This presents a signficiant challenge for a precice deposition of the liquid delivered by an airtanker and complicates an accurate assessment of aircraft-based liquid delivery systems in the popular cup-and-grid method, the measurement grid of which usually has a width of $43-100 \, \text{m}$ \citep{Suter2000,Gu2023}. Hence, a droplet with an initial radius of $300 \, \upmu\text{m}$, which corresponds approximately to the modal droplet size expected from the breakup of ligaments produced by the Kelvin-Helmholtz instability, see Figure \ref{fig:plane-DSD}(b), can be carried beyond the measurement grid by crosswinds with velocities as small as $1-2\, \mathrm{m/s}$.

\begin{figure}[t]
	\centering
	\includegraphics[width=\linewidth]{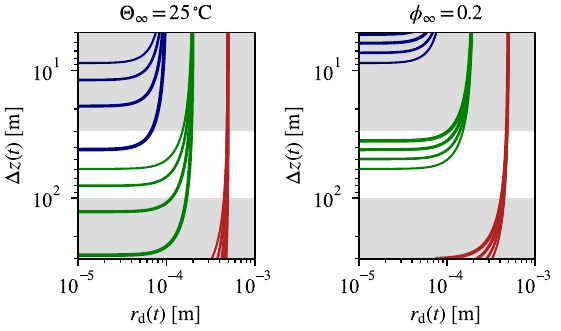}
	\caption{The falling distance $\Delta z$ as a function of the radius $r_\text{d}$ of water droplets falling freely in a quiescent air atmosphere, with an initial droplet temperature of $\Theta_\text{d,0}=25 \, ^\circ\text{C}$ and initial droplet radii of $r_\text{d,0} \in \{100, 200, 400\} \, \upmu\text{m}$. The typical release height of $30-100 \, \text{m}$ in airtanker firefighting is explicitly highlighted. (a) The atmosphere has a temperature of $\Theta_\infty = 25 ^\circ\text{C}$ and relative humidity $\phi_\infty \in \{0.2,0.4,0.6,0.8\}$, with thicker lines corresponding to larger $\phi_\infty$. (b) The atmosphere has a relative humidity of $\phi_\infty = 0.2$ and initial temperature $\Theta_\infty \in \{25,30,35,40\} \, ^\circ \text{C}$, with thicker lines corresponding to larger $\Theta_\infty$. }
	\label{fig:R-z_phi-T}
\end{figure}

\begin{figure}[h]
	\centering
	\includegraphics[width=\linewidth]{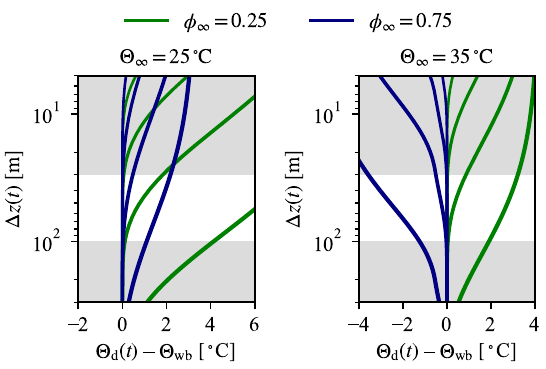}
	\caption{Evolution of the droplet temperature $\Theta_\text{d}$ relative to the wet-bulb temperature $\Theta_\text{wb}$ with respect to the falling distance $\Delta z$ of water droplets falling freely in a quiescent air atmosphere with temperature $\Theta_\infty \in \{ 25, 35\} ^\circ\text{C}$ and relative humidity $\phi_\infty \in \{0.25,0.75\}$. The droplets have an initial temperature of $\Theta_\text{d,0} = 25 \, ^\circ\text{C}$ and initial radii $300 \, \upmu\text{m} \leq r_\text{d,0} \leq 3 \, \text{mm}$, with thicker lines corresponding to larger initial radii. The wet-bulb temperature is calculated by Eq.~\eqref{eq:Twb}. The typical release height of $30-100 \, \text{m}$ in airtanker firefighting is explicitly highlighted.}
	\label{fig:t_Td-real}
\end{figure}

\subsection{Heat and mass transfer}

To better understand the influence of the atmospheric conditions characterized by the ambient temperature and relative humidity, Figure \ref{fig:R-z_phi-T} considers the fall of droplets with an initial radius of $r_\text{d,0} \in \{100, 200, 400\} \, \upmu\text{m}$ in different conditions. A change in relative humidity is observed to have a more significant impact on the droplet evaporation, and thus on the droplet sizes that survive the fall, than a change in ambient temperature. Consequently, although the ambient temperature influences the droplet evaporation, it appears to be the humidity that dominates the mass transfer from the droplet to its surroundings. This is an important finding because tests conducted to evaluate the effectiveness of airtankers do not record the relative humidity on site and studies of airtanker firefighting published to date have not taken the influence of humidity into account \citep{Amorim2011a,Legendre2014,Legendre2024, Sun2024a, Gu2023}.

As expected based on an equilibrium of sensible and latent heat flux, small evaporating droplets quickly assume a liquid temperature equal to the wet-bulb temperature.
Figure \ref{fig:t_Td-real} shows the evolution of the droplet temperature $\Theta_\text{d}$ relative to the wet-bulb temperature $\Theta_\text{wb}$ with respect to the falling distance $\Delta z$ of water droplets falling freely in a quiescent air atmosphere. While droplets with an initial radius of $300 \, \upmu\text{m}$ or less assume the wet-bulb temperature after a fall of only a few meters, millimeter-sized droplets are typically not able to assume the wet-bulb temperature before reaching the ground. Hence, determining the evaporation time of droplets in airtanker firefighting by using a simple relationship such as Eq.~\eqref{eq:tauev} is a rough approximation at best, even if the Sherwood number is based on an \textit{a priori} estimation of the terminal Reynolds number of the droplet.

\section{Evaporated droplet volume}
\label{sec:evpoarated-volume}

The evaporated droplet volume is a key measure of the effectiveness of an aerial water release, as only droplets that can get close to a fire contribute to cooling or extinguishing it, and only droplets that reach the ground contribute to the liquid volume measured during tests used to assess airtanker delivery systems. 
As a result of the primary atomization process, a release of a water volume of $3 \, \text{m}^3$ or more by a large airtanker results in $\mathcal{O}(10^{10})$ individual droplets. It is, thus, reasonable to assume that the evaporation process results in near saturated conditions in the core of the resulting spray, with a relative humidity close to $100 \, \%$ and an air temperature close to the wet-bulb temperature. This corresponds, in fact, to the conditions found in atmospheric clouds \citep{Pruppacher2010}. Hence, two limiting cases for the ambient conditions of the droplets ought to be considered: ambient conditions of the atmosphere at the edge of the spray and near-saturated conditions in the core of the spray. 

\begin{figure*}[h!]
	\centering
	\includegraphics[width=0.495\linewidth]{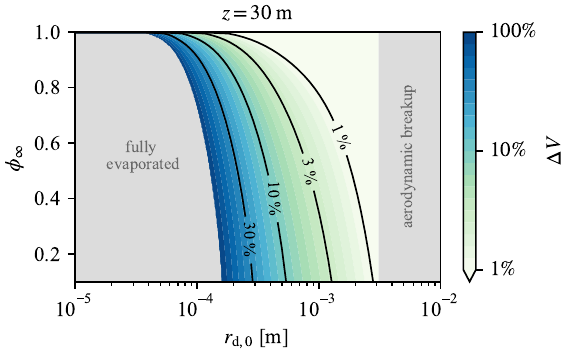}\hfill
	\includegraphics[width=0.495\linewidth]{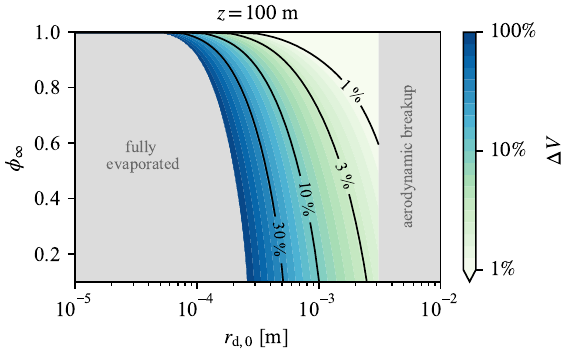}
	\caption{Phase maps of the evaporated volume $\Delta V$ of water droplets falling freely in a quiescent air atmosphere with a temperature of $\Theta_\infty = 25 \, ^\circ\text{C}$ for a release from a height of $z \in \{30,100\} \, \text{m}$, as a function of the initial droplet radius $r_\text{d,0}$ and the relative humidity $\phi_\infty$. The initial droplet temperature is $\Theta_\text{d,0} = 25 \, ^\circ\text{C}$.}
	\label{fig:phasemap_evaporated}
\end{figure*}

Figure \ref{fig:phasemap_evaporated} shows phase maps of the evaporated volume $\Delta V$ of droplets falling freely in a quiescent atmosphere with temperature $\Theta_\infty = 25 \, ^\circ\text{C}$ from heights of $z=30 \, \text{m}$ and $z=100 \, \text{m}$, as a function of the initial droplet radius and the relative humidity. The evaporation and breakup of droplets provide clear lower and upper boundaries for the size of free-falling droplets that reach the ground. As a result, a narrow band of initial droplet radii reaching the ground emerges, which is strongly dependent on the release height and the ambient relative humidity. Conveniently for airtanker firefighting, this band of initial droplet radii largely overlaps with the droplet radii that are expected to be produced by the primary atomization process, although the smallest droplets produced thereby evaporate quickly. The phase maps in Figure \ref{fig:phasemap_evaporated} highlight a clear advantage of millimeter-sized droplets, which lose only a small fraction of their initial volume during an aerial release, whereas droplets with an initial radius $r_\text{d,0} < 1 \, \text{mm}$ are vulnerable to the ambient conditions and release height. 

To better understand the influence of evaporation on airtanker firefighting, Monte Carlo simulations of a larger number of droplets are conducted. These Monte Carlo simulations should be interpreted as a statistical sampling of independent droplet trajectories rather than a direct representation of the total droplet population. The radii of these droplets are initialized with the three gamma distributions identified by \citet{Legendre2024}, see Figure \ref{fig:plane-DSD}(b), defined as \citep{Villermaux2007}
\begin{equation}
	\mathcal{P}(x) = \frac{k^k}{\Gamma(k)} x^{k-1} \text{e}^{-kx}, \label{eq:gamma}
\end{equation}
where $x=r_\text{d,0}/\bar{r}_\text{d,0}$, $\bar{r}_\text{d,0}$ is the mean initial radius, $k$ is the order of the gamma distribution, and $\Gamma(k) = \int_0^\infty x^{k-1} \text{e}^{-x} \, \text{d}x$ is the gamma function. Following the work of \citet{Legendre2024}, three droplet production mechanisms related to the prevailing high-shear Kelvin-Helmholtz instability are considered: ligament breakup ($k=4.7$, $\bar{r}_\text{d,0} = 339 \, \upmu\text{m}$), bag rim breakup ($k=4$, $\bar{r}_\text{d,0} = 669 \, \upmu\text{m}$), and bag film breakup ($k=11$, $\bar{r}_\text{d,0} = 319 \, \upmu\text{m}$). The loss of liquid volume $\Delta V$ of $10^4$ water droplets (per distribution) falling from a height of $z \in \{30,100\}\, \text{m}$ in different ambient conditions, with and without crosswinds, is listed in Table \ref{table:dsd-evaporation}. In general, since large droplets evaporate slowly and hold a disproportionate share of the liquid volume ($V_\text{d} \propto r_\text{d}^3$), the droplets produced by bag rim breakup ($k=4$, $\bar{r}_\text{d,0} = 669 \, \upmu\text{m}$) lose the least amount of liquid volume due to the larger droplet mean size. The relative humidity, again, emerges as an important ambient property, with the ambient temperature having a comparably small influence. It is important to remember that the relative humidity is typically not recorded during tests, revealing an important shortcoming of test protocols for airtanker delivery systems. The release height has also a significant influence on the liquid volume that reaches the ground, because a fall from larger heights allows the droplets more time to evaporate. Consequently, in a quiescent atmosphere,  a release from $z=100 \, \text{m}$ results in a volume loss $2.5-3$ times larger than a release from $z=30 \, \text{m}$.

\begin{table*}[h]
	\centering
	\caption{Lost volume $\Delta V$ of water droplets belonging to the droplet size distributions initialized with the three gamma distributions shown in Figure \ref{fig:plane-DSD}(b) falling in an air atmosphere under different release conditions. The initial droplet temperature is $\Theta_\text{d,0} = 25 \, ^\circ\text{C}$.}\label{table:dsd-evaporation}
	\begin{tabular}{cccc|ccccc}
		 & & & & $\Delta V$ & $\Delta V$ & $\Delta V$\\
		$z$ & $u_\text{cross}$ & $\Theta_\infty$ & $\phi_\infty$ & $k=4.7$, $\bar{r}_\text{d,0} = 339 \, \upmu\text{m}$ & $k=4$, $\bar{r}_\text{d,0} = 669 \, \upmu\text{m}$ & $k=11$, $\bar{r}_\text{d,0} = 319 \, \upmu\text{m}$ \\ 
		 \hline 
		 \multirow{5}{*}{$30 \, \text{m}$} & \multirow{3}{*}{$0 \, \text{m/s}$} & $25 \, ^\circ\text{C}$ & $0.25$  & $11.5 \, \%$ & $\phantom{0}3.7 \, \%$ & $16.8 \, \%$ \\
		 & & $25 \, ^\circ\text{C}$ & $0.75$ & $\phantom{0}3.6 \, \%$ & $\phantom{0}1.2 \, \%$ & $\phantom{0}5.1 \, \%$ \\
		 & & $35 \, ^\circ\text{C}$ & $0.25$ & $14.6 \, \%$ & $\phantom{0}4.1 \, \%$ & $22.0 \, \%$ \\
		 & & $35 \, ^\circ\text{C}$ & $0.75$ & $\phantom{0}3.2 \, \%$ & $\phantom{0}0.1 \, \%$ & $\phantom{0}5.4 \, \%$ \\
		 \cline{2-7}
		 & $2 \, \text{m/s}$ & $25 \, ^\circ\text{C}$ & $0.25$ & $11.6 \, \%$ & $\phantom{0}3.7 \, \%$ & $16.9 \, \%$ \\
		 & $5 \, \text{m/s}$ & $25 \, ^\circ\text{C}$ & $0.25$ & $43.8 \, \%$ & $\phantom{0}6.1 \, \%$ & $79.8 \, \%$ \\
		 \hline 
		 \multirow{5}{*}{$100 \, \text{m}$} & \multirow{3}{*}{$0 \, \text{m/s}$} & $25 \, ^\circ\text{C}$ & $0.25$ & $30.6 \, \%$ & $\phantom{0}9.4 \, \%$ & $46.0 \, \%$ \\
		 & & $25 \, ^\circ\text{C}$ & $0.75$ & $10.1 \, \%$ & $\phantom{0}2.9 \, \%$ & $15.1 \, \%$ \\
		 & & $35 \, ^\circ\text{C}$ & $0.25$ & $39.2 \, \%$ & $11.8 \, \%$ & $58.6 \, \%$ \\
		 & & $35 \, ^\circ\text{C}$ & $0.75$ & $12.6 \, \%$ & $\phantom{0}2.2 \, \%$ & $20.2 \, \%$ \\
		  \cline{2-7}
		 & $2 \, \text{m/s}$ & $25 \, ^\circ\text{C}$ & $0.25$ & $84.8 \, \%$ & $21.6 \, \%$ & $99.7 \, \%$ \\
		 & $5 \, \text{m/s}$ & $25 \, ^\circ\text{C}$ & $0.25$ & $100 \, \%$ & $99.5 \, \%$ & $100 \, \%$ \\
		 \hline 
	\end{tabular}
\end{table*}

Including a crosswind highlights the importance of a careful consideration of wind speed and direction, especially during field tests. The results presented in Table \ref{table:dsd-evaporation} assume a measurement area with a generous total width of $100 \, \text{m}$, with the airplane releasing water along the centerline of that measurement area. For a crosswind with a velocity of $u_\text{cross} = 5 \, \text{m/s}$, all droplets with a time of flight larger than $10 \, \text{s}$ either evaporate fully or land outside the measurement area. Based on Figure \ref{fig:falltime}, for a release height of $z = 30 \, \text{m}$ this means only droplets with an initial radius $r_\text{d,0} \gtrsim 500 \, \upmu\text{m}$ contribute to the retained liquid volume, often called the \textit{ground fraction}, whereas almost the entire liquid volume evaporates or is carried beyond the measurement area if released from a height of $z = 100 \, \text{m}$.

Considering the cases in which the droplets fall from $z = 30 \, \text{m}$ and the ambient air has a relative humidity of $\phi_\infty = 0.75$, it is interesting to note that less liquid volume evaporates for an ambient temperature of $\Theta_\infty = 35 \, ^\circ\text{C}$ than for $\Theta_\infty = 25 \, ^\circ\text{C}$. This counterintuitive observation can be explained by the initial droplet temperature of $\Theta_\text{d,0} = 25 \, ^\circ\text{C}$ in comparison to the wet-bulb temperature $\Theta_\text{wb}$ at these conditions. 
At $\Theta_\infty = 25 \, ^\circ\text{C}$ the wet-bulb temperature is $\Theta_\text{wb} = 21.6 \, ^\circ\text{C}$, meaning that the droplets evaporate right from the beginning and cool to the wet-bulb temperature. In contrast, at $\Theta_\infty = 35 \, ^\circ\text{C}$ the wet-bulb temperature is  $\Theta_\text{wb} = 31.1 \, ^\circ\text{C}$, meaning that the droplets initially gain volume due to condensation as they heat up to the wet-bulb temperature. While this initial condensation phase is negligible for small droplets, droplets with an initial radius of $r_\text{d,0} \gtrsim 400 \, \upmu\text{m}$ arrive on the ground with a larger radius for $\Theta_\infty = 35 \, ^\circ\text{C}$ than for $\Theta_\infty = 25 \, ^\circ\text{C}$ under the studied conditions. 

The results discussed in this study correspond to an ideal exposure of all water droplets to the atmosphere, which corresponds approximately to the condition prevailing at the edge of the liquid spray produced by an aerial water release. In contrast, in the core of the spray the evaporation process may lead to near saturated conditions, considerably reducing the evaporation rate. As a consequence, less than $1\, \%$ of the droplet volume evaporates and, hence, droplets with a larger range of sizes may reach the ground. Nevertheless, a lower limit of the initial size of droplets surviving the fall exists even under near saturated conditions, as observed in Figure \ref{fig:phasemap_evaporated}. In reality, however, the complexity of the underlying fluid dynamics likely promotes extensive mixing and results in non-uniform conditions that are not fully understood to date.

\section{Water volume lost in a field test}
\label{sec:cases}

\begin{figure*}[t]
	\includegraphics[width=0.95\linewidth]{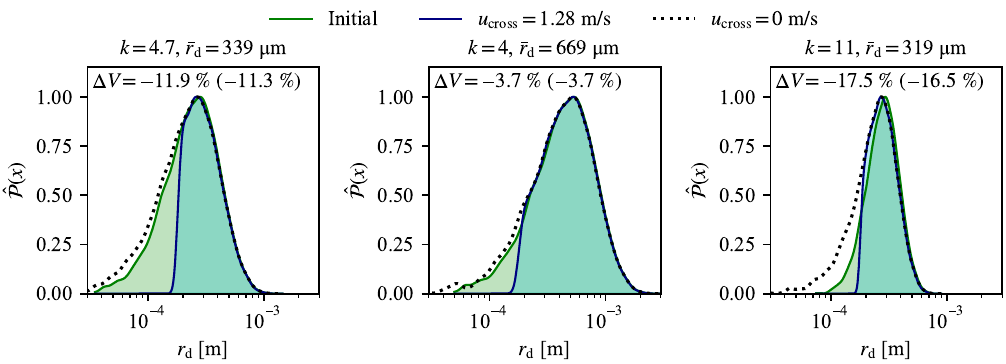}
	\caption{Initial and final droplet size distributions $\hat{\mathcal{P}}(x)$, normalized by their modal values, of a simulation of $10^4$ water droplets (per distribution), falling in an air atmosphere from a height of  $z = 58 \, \text{m}$ over a measurement area with a total width of $100 \, \text{m}$. The lost droplet volume $\Delta V$ in the presence of a crosswind with velocity $u_\text{cross} = 1.28 \, \text{m/s}$ is indicated explicitly, with the corresponding value for a quiescent atmosphere ($u_\text{cross} = 0$) given in parentheses. The droplet radii are initialized with the three representative gamma distributions shown in Figure \ref{fig:plane-DSD}(b).}
	\label{fig:dsd_Gua}
\end{figure*}

A field test with an AG600 amphibious firefighting aircraft reported by \citet{Gu2023} is considered, which took place on 12 April 2022 in the Hubei Province of China. Overall six test where conducted on this day, under similar conditions and yielding similar results. Although no temperature and relative humidity are reported by \citet{Gu2023}, based on historical weather data, the ambient conditions during the test are estimated as $\Theta_\infty = 20 \, ^\circ\text{C}$ and $\phi_\infty=0.5$.  
In the particular test considered here, the aircraft released $6 \, \text{m}^3$ of water from a height of $z = 58 \, \text{m}$, into an atmosphere with a wind of $1.3 \, \text{m/s}$ at an angle of $260^\circ$ with respect to the flight direction (i.e.~with a crosswind component of $u_\text{cross} = 1.28 \, \text{m/s}$). Using the cup-and-grid method with a grid of $300 \, \text{m} \times 100 \, \text{m}$, a ground fraction (i.e.~the share of liquid that arrived on the ground) of $41.35 \, \%$ was measured. \citet{Gu2023} acknowledged that parts of the liquid dropped outside the measurement grid and was included in the ground fraction by approximating its contribution by linear interpolation. Hence, almost $60 \, \%$ of the released water was lost, presumably due to evaporation or entrainment in the aircraft wake.

Conducting Monte Carlo simulations of $10^4$ droplets initialized with the three representative initial droplet size distributions introduced in Section \ref{sec:evpoarated-volume}, Figure \ref{fig:dsd_Gua} presents the corresponding initial and final droplet size distributions, both in the presence and absence of the prevailing crosswind. For clarity and to enable a direct comparison of their shapes, all distributions are normalized with respect to their modal values. Without a crosswind, meaning that any liquid lost is due to evaporation, the distributions shift to smaller sizes as a result of evaporation. Even though this change is qualitatively small, the resulting reduction in volume is considerable, with approximately $6-30 \, \%$ of the total liquid volume evaporating during its fall. Introducing the crosswind of $u_\text{cross} = 1.28 \, \text{m/s}$, the size distributions of droplets reaching the ground are sharply cut for $r_\text{d} \lesssim 150 \, \upmu \text{m}$, which corresponds to an initial droplet size of $r_\text{d,0} \lesssim 220 \, \upmu \text{m}$. Droplets with an initial radius below this threshold have a time of flight larger than the $39 \, \text{s}$ required for the crosswind to carry the droplets for $50 \, \text{m}$ outside the measurement area. However, the liquid volume lost as a result of wind drift is very small, as the total liquid volume lost, $\Delta V$, is almost unchanged compared to the results obtained in a quiescent atmosphere. Hence, the droplets carried outside the measurement area by the crosswind do not contribute much to the total volume of droplets. A sufficiently wide measurement area with respect to the release height can, thus, reduce the impact of crosswinds on the deposited liquid volume considerably. 

Since evaporation does not account for the reported loss of liquid volume during the test of approximately $59 \, \%$ even under the favorable conditions for evaporation considered here, a substantial number of droplets was likely carried beyond the measurement area by other aerodynamic effects, such as entrainment of the droplets in the aircraft wake. This further underlines the necessity of considering the full velocity field produced by airtankers and highlights the limitations of this study.

\section{Conclusions}
\label{sec:conclusions}

This study has, for the first time, systematically investigated the dynamics of water droplets in airtanker firefighting. While previous research focused on the ground patterns from tests in which airtankers release water or retardant over a dedicated measurement area \citep{Legendre2014,Gu2023,Amorim2011,Sun2024a,Wheatley2023}, the fate of the billions of individual droplets produced by the rapid atomization process that follows the release of the liquid from an airtanker has not yet been considered in detail.

A tailored model for the motion as well as heat and mass transfer of an isolated droplet in an air atmosphere has been presented. With this model, the behavior of a single droplet under different atmospheric and operational conditions has been studied. The presented results highlight the central role of the droplet size in determining the amount of water delivered by airtanker firefighting operations and the release height has been found to be the dominant operational factor influencing both droplet evaporation and wind drift. Because droplets may travel long distances through warm and dry air before reaching the ground or canopy, even a modest increase in release height significantly increases the time of flight of droplets and, consequently, the loss of liquid volume. With respect to the ambient atmospheric conditions, the presented results demonstrate that humidity is more influential than air temperature; low relative humidity strongly accelerates evaporation, particularly for droplets with radii below one millimeter. These small droplets are also more susceptible to displacement by crosswinds, leading to substantial drift away from the intended target area. In summary, only droplets in a narrow range of initial radii, $150 \, \upmu\text{m} \lesssim r_\text{d,0} \lesssim 3 \, \text{mm}$, can reach the ground as a result of a water release from an airtanker, with smaller droplets fully evaporating before reaching the ground or being displaced considerably by wind drift, and larger droplets are broken up by secondary atomization.

Although this study has highlighted a number of important mechanisms that play a key role in the efficacy of water delivery by an airtanker, its underpinning simplifications (e.g.~no air flow, no change of ambient conditions as a result of evaporation) and inherent limitations (e.g.~single droplet analysis) mean that it can only serve as a starting point for a more comprehensive analysis of water release from airtankers. By isolating droplet-scale physics, the present study provides reference behavior and bounding estimates to underpin any future spray-resolved modeling of airtanker water release. The combined effects of rapid evaporation and wind drift observed in this study highlight the operational need to produce larger droplets, which retain their volume longer and follow more stable trajectories. Hence, operational strategies that minimize flight time, i.e.~water release from a lower height, and favor the production of larger droplets ought to be prioritized. For instance, the most unstable mode of the Kelvin-Helmholtz instability shifts to longer wavelengths for smaller velocity differences \citep{Drazin2002}, which suggests that minimizing the airspeed at which water or retardants are released can promote the formation of larger liquid structures. However, irrespective of airspeeds, the droplet size distribution resulting from the primary atomization process following a water release from an airtanker remains essentially unknown \citep{Legendre2024,Rouaix2023,Sun2024a}, preventing a reliable quantification of the droplet dynamics and, consequently, the volume lost to evaporation, wind drift, and entrainment. A key to achieving reliable predictions of airtanker firefighting will, thus, be dedicated future studies of the primary atomization process and the formulation of a tailored spray function, as already pointed out by \citet{Legendre2024}, leveraging high-fidelity numerical simulations, laboratory experiments, as well as carefully controlled field tests.


\end{document}